\documentclass[twocolumn,showpacs,preprintnumbers,amsmath,amssymb,aps]{revtex4}
\usepackage{graphicx}% Include figure files
\usepackage{dcolumn}% Align table columns on decimal point
\usepackage{bm}% bold math

\begin{document}

\preprint{APS/123-QED}

\title{Parallel tale of seniority isomers in $^{130}$Cd and $^{206}$Hg: Testing the robustness of magic numbers}% Force line breaks with \\
\author{Bhoomika Maheshwari}
\email{bhoomika.physics@gmail.com}
\author{Hasan Abu Kassim}
\author{Norhasliza Yusof}
\affiliation{Department of Physics, Faculty of Science, University of Malaya, 50603 Kuala Lumpur, Malaysia. \\
\\
Center of Theoretical Physics, Department of Physics, University of Malaya,50603 Kuala Lumpur, Malaysia.}
\author{Ashok Kumar Jain}%
\affiliation{Amity Institute of Nuclear Science and Technology, Amity University Uttar Pradesh, 201313 Noida, India.}

\date{\today}% It is always \today, today,

\begin{abstract}
\noindent\textbf{Background:} The neutron-rich nuclei $^{130}$Cd and $^{206}$Hg, so important in the astrophysical processes, may also be useful in tracking the evolution of nuclear shell gaps as one traverses the neutron-rich region. The high spin $8^+$ isomer in $^{130}$Cd and the ${10}^+$ isomer in $^{206}$Hg turn out to be the lamp posts, which may shed light on the shell gaps and validity of the seniority scheme in the neutron-rich systems. \\
\textbf{Purpose:} We explore the robustness of the N=82 and N=126 magic numbers in the neutron-rich $^{130}$Cd and $^{206}$Hg nuclides, respectively. A parallel between the two nuclides in terms of the high-spin isomers allows us to investigate these ‘waiting-point’ nuclei, which have limited experimental data, by using the concept of seniority as the stepping stone. \\
\textbf{Method:} In this paper, we report large scale shell model calculations by using the available realistic effective interactions derived from the Charge Dependent Bonn potential through the renormalized G matrix. We also explore if any change in the interaction is also required to consistently explain both the level structures as well the B(E2) values.\\
\textbf{Results:} The shell model calculations with the standard “$jj45pna$” and “$CWG$” interactions explain the B(E2) values quite well in $^{130}$Cd and $^{206}$Hg, respectively. However, a consistent description of the level schemes as well as the B(E2)s requires a modification of the interaction in terms of the two-body matrix elements of $\Pi g_{9/2}^{-2}$ and $\Pi h_{11/2}^{-2}$, for $^{130}$Cd and $^{206}$Hg, respectively. A structural similarity between the $8^+$ isomer in $^{130}$Cd and the ${10}^+$ isomer in $^{206}$Hg is noticed due to goodness of seniority. They are found to possess a maximally aligned, seniority $v=2$ configuration from their respective intruder orbits.\\
\textbf{Conclusion:} The two isomeric states turn out to be pure seniority meta-stable states, where the highest $j$ value involved is $j=9/2$ and $11/2$. No shell quenching seems to be needed for the seniority isomers in these nuclei. Therefore, N=82 and 126 appear to be very robust magic numbers even in the neutron-rich region.

% Valid PACS numbers may be entered using the \verb+\pacs{#1}+ command.
\end{abstract}

\maketitle

\section{\label{sec:level1}Introduction}

Existence of a large number of isomers across the nuclear landscape enables a unique insight into the nuclear structure properties of different mass regions~\cite{jain2015}. Besides, many of these isomers are beginning to have many multidisciplinary applications which make them a hot and trending topic across the globe. Be it the nuclear clock or the dream of gamma ray lasers, isomers appear to be of great significance~\cite{wense2016, wense2018, seiferle2019}. Even as more and more data are becoming available, a host of new questions related to the decay properties of isomers are beginning to arise ~\cite{jain2015, walker1999, walker2005, walker2007, dracoulis2013, dracoulis2016, walker2017}. A simple dictum is, “the larger the decay gets hindered, longer the isomer may live”. Based on the hindrance mechanisms, isomers may be classified into five categories: spin isomers, K isomers, shape isomers, fission isomers and seniority isomers. The “Atlas of Nuclear Isomers” has allowed us to identify many systematic features in these isomers~\cite{jain2015}. Some of these features have been understood by using the generalized seniority scheme for identical nucleons~\cite{maheshwari2016, maheshwari20161, jain2017, jain20171, maheshwari2017}. As a result, we could establish a new set of selection rules and also a new kind of seniority isomers, which decay by odd tensor electric transitions~\cite{maheshwari2016}. 

Seniority isomers arise in semi-magic nuclei due to the specific selection rules related to the seniority quantum number~\cite{casten1990, heyde1990, talmi1993, isacker2011, isacker2014, qian2018, talmi2018}. Sn isotopes have served as the most common text book examples of the seniority scheme ~\cite{casten1990, heyde1990, talmi1993}. The neutron-rich region around $^{132}$Sn have also attracted much attention both theoretically and experimentally due to its importance in nuclear physics and astrophysics~\cite{jain20171, simpson2014, maheshwari2015, dillmann2003, jungclaus2007}. For example, the large scale shell model (LSSM) calculations of the $6^+$ isomers in nuclei beyond $^{132}$Sn suggest a marginal modification of the realistic effective interaction to explain the seniority mixing in $^{136}$Sn ~\cite{simpson2014, maheshwari2015}. However, the usage of generalized seniority resolved the observed behavior of the highly neutron-rich Sn isotopes beyond $N=82$~\cite{jain20171} and also reinforced the need to modify the two-body matrix elements (TBME) as well the single particle energy of the $i_{13/2}$ orbit.

In a similar spirit, we study the $8^+$ isomer in $^{130}$Cd ($Z=48$ and $N=82$) arising from seniority $v=2$, $g_{9/2}^{-2}$ (two proton holes) configuration. We further draw a parallel with the ${10}^+$ isomer in $^{206}$Hg ($Z=80$ and $N=126$) arising from seniority $v=2$, $h_{11/2}^{-2}$ (two proton holes) configuration. $^{130}$Cd is a classical r-process $N=82$ waiting-point nucleus with very limited data ~\cite{kratz1986, kratz1988, pfeiffer2001, hannawald2001} until a recent experiment reported by Dillmann $et$ $al.$~\cite{dillmann2003}. They highlighted the importance of shell quenching in Cd isotopes and highlighted the need of further confirmation by new measurements in $^{129}$Cd and $^{131,132}$Cd. Such a shell quenching may arise due to diffusion of nuclear potential in very neutron-rich nuclei resulting in significant modifications of the single particle energies and the related shell gaps. The usual shell gaps in the neutron-rich regions may be less pronounced or, transient compared to those in the stable nuclei.
 
On the other hand, the tensor nature of nuclear force may also lead to a reordering of the shell gaps in very proton/neutron-rich nuclei. Jungclaus $et$ $al.$~\cite{jungclaus2007} observed the excited states in $^{130}$Cd for the first time in 2007, and the $8^+$ isomer (whose spin-parity assignment is tentative till date) appeared to restore the $N=82$ shell gap in this nucleus. So the shell quenching at $N=82$ in $^{130}$Cd is still questionable and leads to many contradicting explanations in literature. Unfortunately, highly neutron-rich $N=82$ waiting-point nuclei are still out of reach experimentally. So an answer to the question of whether and how far below $^{132}$Sn, an erosion of the $N=82$ shell gap occurs, will need to wait till the lighter $N=82$ waiting-point nuclei are reached by the future radioactive beam facilities. Besides, Dunlop $et$ $al.$~\cite{dunlop2016} and Jungclaus $et$ $al.$~\cite{jungclaus2016} recently studied the beta-decay properties of $^{130}$Cd and highlighted their crucial role in nuclear structural and astrophysical r-process calculations. In view of this, the single-particle structure and the applicability of the seniority and generalized seniority scheme may prove to be a useful approach to investigate this region and make some useful predictions.

As already stressed, it is also interesting to compare the $8^+$ isomer in $^{130}$Cd with the ${10}^+$ isomer in $^{206}$Hg near the other doubly-magic cornerstone nucleus $^{208}$Pb with $N=126$, particularly for probing the interactions used in the shell model. A change in the nature of the interaction used in shell model can significantly alter the results. Although many new isotopes/isotones around $N=126$ have been accessed in recent times ~\cite{alvarez2010, morales2011, kurcewicz2012, pietralla2014}, there remain many gaps in this region of nuclear chart. A recent isomeric study~\cite{lalovic2018} in the vicinity of $^{208}$Pb covered the isomers in $^{198,200,202,206}$Pb and $^{206}$Hg by using the relativistic projectile fragmentation and used the shell model calculations with the Kuo-Herling interaction to understand them~\cite{mcgrory1975}. It would be nice to check other available realistic effective interactions in this region to understand the occurrence of the ${10}^+$ isomer (with a tentative spin-parity assignment) in $^{206}$Hg.
 
The chosen pair of nuclei, $^{130}$Cd and $^{206}$Hg, are known to have two proton holes, although in different valence spaces around different magic configurations. A comparative study of these two nuclei should reveal the evolution of nuclear structure in neutron-rich nuclei around $Z=50$ (with $N=82$ closed shell) and $Z=82$ region (with $N=126$ closed shell). This comparison is of particular interest due to seniority, which may dictate a similar behavior of the isomers and the trends of their reduced electric transition probabilities B(EL) in different valence spaces involving different sets of orbits. We have investigated the shell evolution and realistic effective interactions used in shell model via LSSM calculations and also verified the role of seniority in their level schemes, B(E2) trends, effective single particle energies and the occurrence of meta-stable states i.e. isomers. In the present paper, the pure seniority scheme explains the origin of isomers in the neutron-rich nuclei. This is largely due to the location of the high-j orbits and the 2-hole nature of the nuclei treated in this paper. However, the description of generalized seniority may be needed to explain the neutron-rich lighter mass waiting-point nuclei (with $N=82$ and $N=126$ closed shells), as we found in neutron-rich Sn isotopes~\cite{jain20171}. This situation can be confirmed in future experiments. 

This paper has been divided into four parts; section 2 presents the details of shell model calculations, valence spaces, single-particle orbits, their energies and the effective interactions. Section 3 presents a comparison of the calculated results with the available experimental data and also discusses the nuclear structural properties, like level energies and B(E2) values, effective single particle energies etc. Section 4 concludes the present paper.

\begin{figure*}[!ht]
\includegraphics[width=16cm,height=14cm]{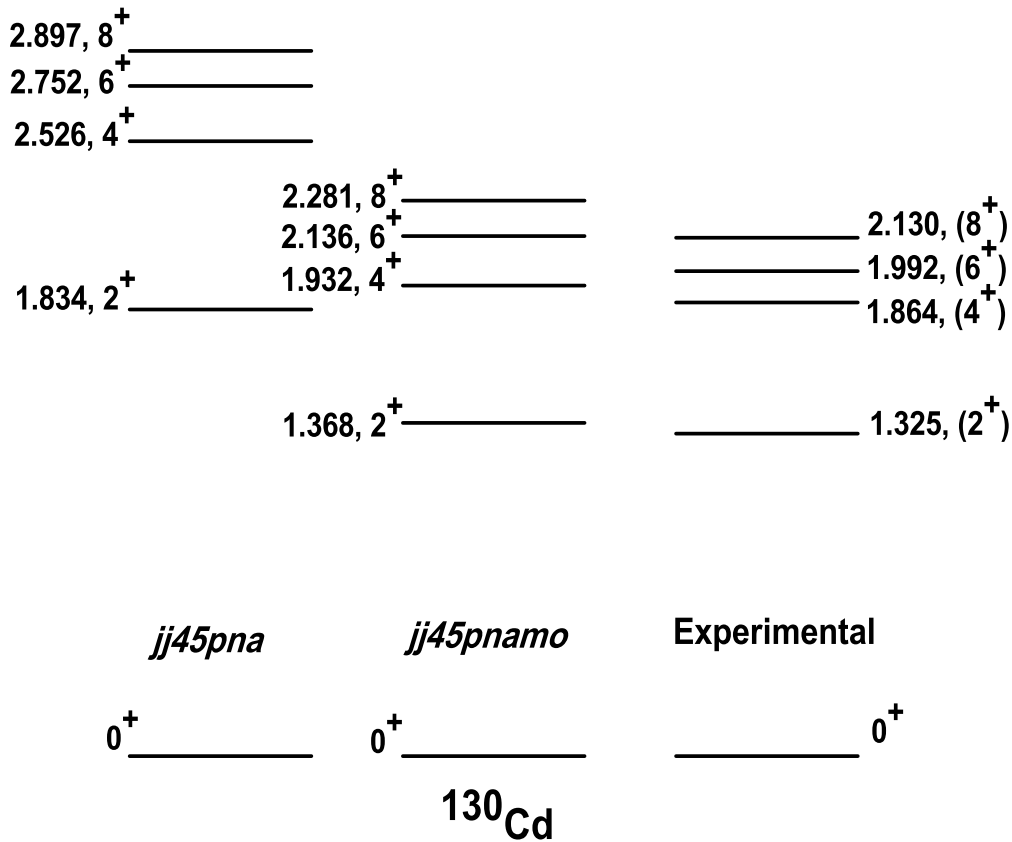}% Here is how to import EPS art
\caption{\label{fig:fig1} The level schemes for $0^+$ to $8^+$ yrast states in $^{130}$Cd calculated with the “$jj45pna$” and modified “$jj45pnamo$” interactions, along with the experimental data~\cite{ensdf}. Note that the spin-parity assignments as well as the location of $6^+$ and $8^+$ states in the experimental data are tentative; it is, however, supported by the LSSM calculations.}
\end{figure*}

\begin{figure*}[!ht]
\includegraphics[width=16cm,height=12cm]{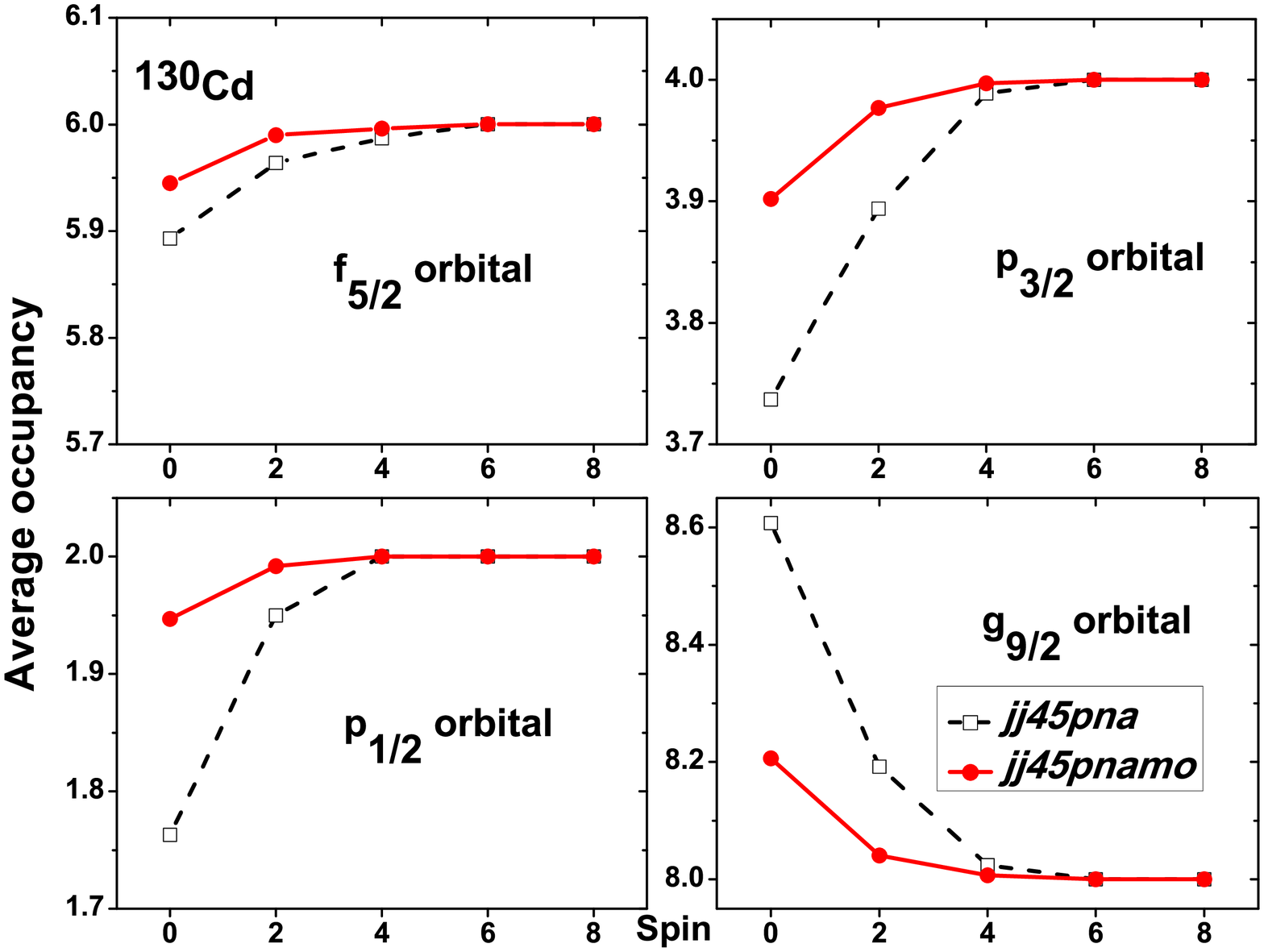}% Here is how to import EPS art
\caption{\label{fig:fig2}(Color online) The calculated average occupancies of the valence protons in $f_{5/2}$, $p_{3/2}$, $p_{1/2}$ and $g_{9/2}$ orbits from the $0^+$ to $8^+$ spins for $^{130}$Cd with “$jj45pna$” and modified “$jj45pnamo$” interactions. These occupancies highlight the dominance of $\Pi g_{9/2}^{-2}$ configuration. }
\end{figure*}
\begin{figure}[!ht]
\includegraphics[width=9.5cm,height=9cm]{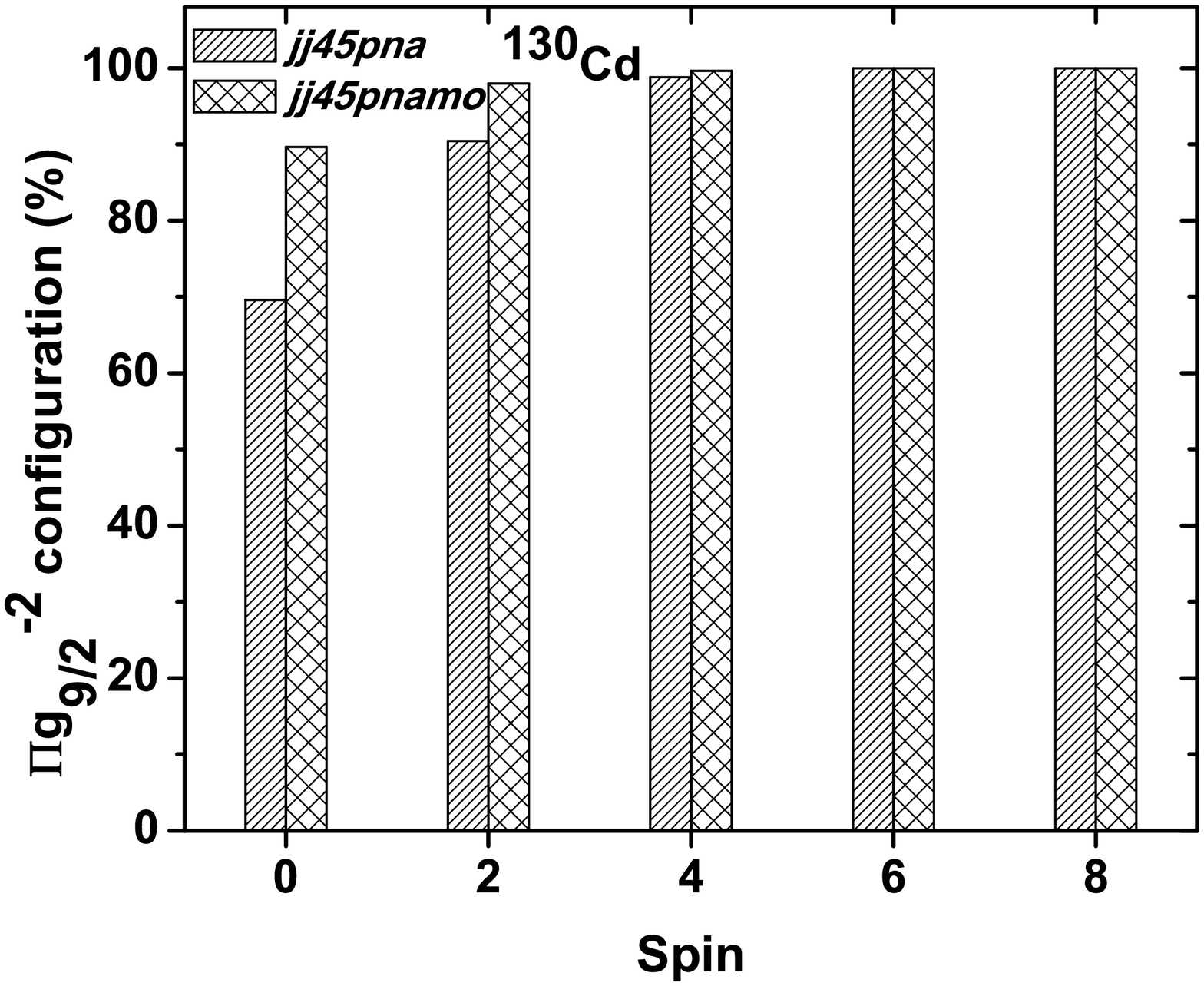}% Here is how to import EPS art
\caption{\label{fig:fig3}(Color online) The calculated $\Pi g_{9/2}^{-2}$ configuration as a percentage of the wave function of the valence protons in the $0^+$ to $8^+$ states for $^{130}$Cd with original “$jj45pna$” and modified “$jj45pnamo$” interactions. }
\end{figure}
\begin{figure}[!ht]
\includegraphics[width=9.5cm,height=9cm]{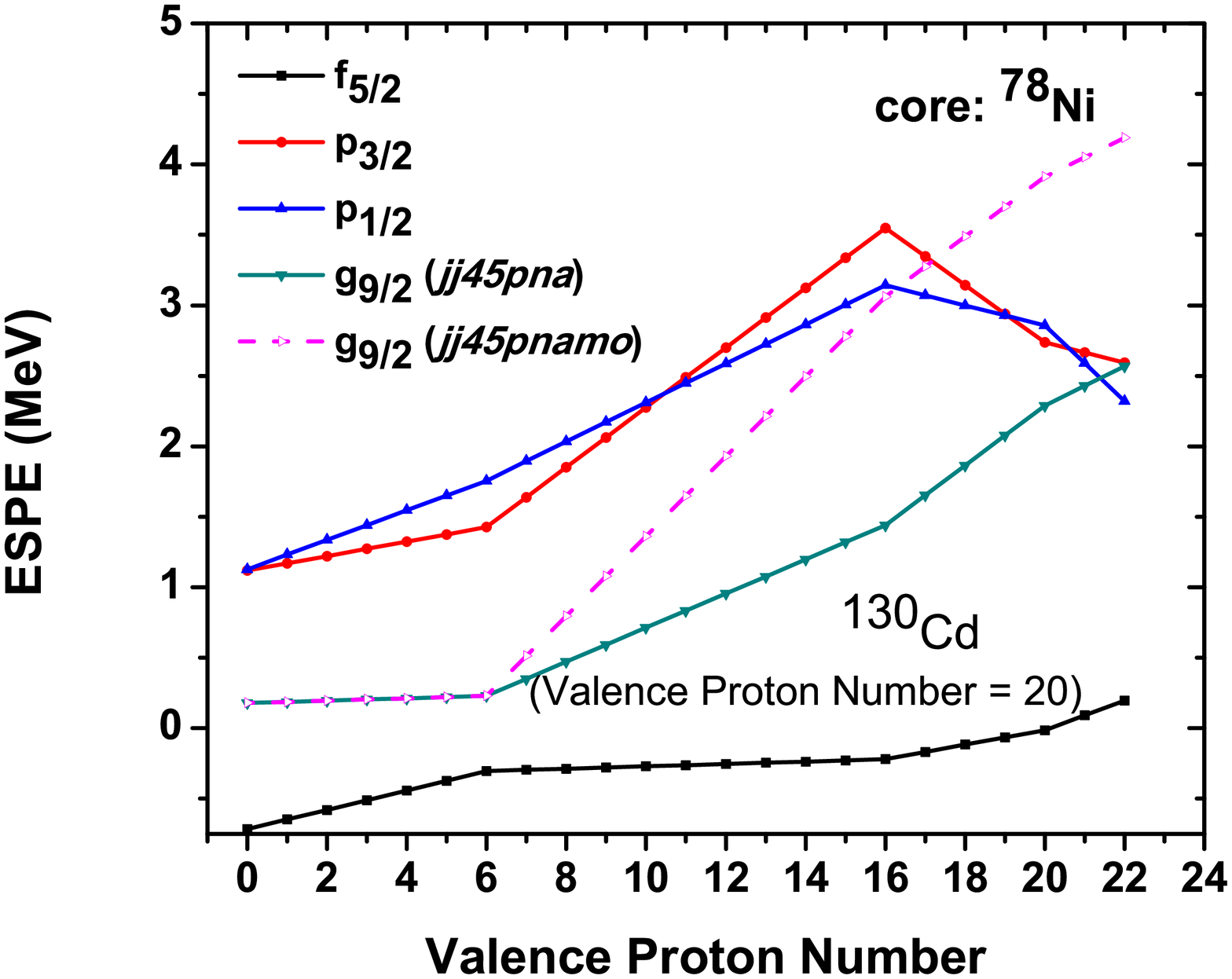}% Here is how to import EPS art
\caption{\label{fig:fig4}(Color online) The variation of ESPE for the active proton orbits $f_{5/2}$, $p_{3/2}$, $p_{1/2}$ and $g_{9/2}$. The ESPE of $g_{9/2}$ have been shown both with original “$jj45pna$” and modified “$jj45pnamo$” interactions. One can observe the large monopole shift at valence proton number $=22$ for $^{130}$Cd in terms of $\Pi g_{9/2}$ after a marginal change of 100 keV in the diagonal $\Pi g_{9/2}^{-2}$ TBME in both proton-proton and Coulomb parts of the interaction.}
\end{figure}

\section{\label{sec:level2}Shell model calculations, valence spaces and the effective interactions}

We have used the NUSHELL code of Brown $et$ $al.$~\cite{brown2007} to carry out the LSSM calculations. The effective interactions for both $^{130}$Cd and $^{206}$Hg were derived from the Charge Dependent Bonn potential through the renormalized G matrix method of Hjorth-Jensen; all the details are given in ref.~\cite{jensen1995, holt2000, engeland2000, brown2005}. The interaction “$jj45pna$” used for $^{130}$Cd assumes the doubly closed $Z=28$, $N=50$, $^{78}$Ni as the core. The active valence space ($Z=28-50$) consists of four proton orbits $1f_{5/2}, 2p_{3/2}, 2p_{1/2}$ and $1g_{9/2}$ having their single-particle energies as $-0.7166, 1.1184, 1.1262$ and $0.1785$ MeV, respectively. This means that the proton excitations across the $Z=50$ shell gap and the neutron excitations across the $N=82$ shell gap are not included. 

On the other hand, the interaction “$CWG$” used for $^{206}$Hg assumes $Z=50$, $N=82$, $^{132}$Sn as the inert core. The valence space $(Z=50-82)$ consists of five proton orbits $1g_{7/2}, 2d_{5/2}, 2d_{3/2}, 3s_{1/2}$ and $1h_{11/2}$ with their single-particle energies as $-9.6630, -8.7010, -6.9950, -7.3230$ and $-6.8700$ MeV, respectively. The proton and neutron excitations across $Z=82$ and $N=126$ respectively, have not been considered. The harmonic oscillator potential is chosen with an oscillator parameter $\hbar \omega= 45 A^{-1/3}$ $- $ $25 A^{-2/3}$. The effective charges of protons and neutrons are taken as $1.5$ and $0.5$, respectively, for calculating the B(E2) values in both the cases. We present the detailed results from $0^+$ to $8^+$ ($0^+$ to ${10}^+$) states in $^{130}$Cd ($^{206}$Hg) and the resulting B(E2) values with both original and modified interactions in the following section, along with the details of modifications in the interactions. A comparison with the available experimental data has also been made and important conclusions are drawn.

\section{\label{sec:level3}Results and Discussion}

\subsection{$^{130}$Cd}

Figure 1 compares the calculated level energies for $^{130}$Cd with the known experimental data. The experimental ordering of the $4^+$, $6^+$ and $8^+$ states is not firmly established because their energies are close to each other. The present spin-parity assignment in $^{130}$Cd ($Z=48$, $N=82$) is based on the comparative level scheme of $^{98}$Cd ($Z=48$ and $N=50$)~\cite{ensdf}, where neutrons are filled up to the magic number $50$. We find that the energies of the $2^+$ to $8^+$ states have been overestimated by the LSSM calculations. However, the calculated ratio $E(4^+)/E(2^+)=1.377$ matches quite well with the experimental value of $1.406$. This confirms the non-collective nature of the levels. These levels seem to arise from pure $\Pi g_{9/2}^{-2}$ (seniority $v=2$) single particle configuration. The observed pattern of the levels is typical of seniority excitations where the energy gap between levels decreases on increasing spin for a given-j. The E2 decay energy for the $8^+$ state is the lowest and gives rise to a long half-life making it an isomeric state. The experimental and calculated separation of $8^+$ and $6^+$ level energies are of similar order as $138$ and $145$ keV, respectively. The experimental [38] and calculated B(E2) values come out to be $50$ and $46$ $e^2fm^4$, respectively. These calculated values correspond to the original “$jj45pna$” interaction.
 
We have then modified the interaction by increasing the two-body diagonal matrix elements of $\Pi g_{9/2}^{-2}$ by only $100$ keV in both proton-proton and Coulomb parts of the interaction. This means that only five TBME $\langle \Pi g_{9/2}^{-2} \mid V \mid  \Pi g_{9/2}^{-2} \rangle$, corresponding to $0^+$, $2^+$, $4^+$, $6^+$ and $8^+$ states have been modified, where V represents a two-body interaction. We call this modified interaction as “$jj45pnamo$” and it leads to a better matching of level energies, as shown in Fig.~\ref{fig:fig1}; this small change shifts the nature of the $0^+$ to $8^+$ states towards pure seniority $v=2$, $\Pi g_{9/2}^{-2}$ configuration.

%%%%%%%%%%%%%%%%%%%%%%%%%%%%%%%%%%%%%%%%%%%%%%%%%%%%%%%%%%%%%%%%%%%%%%%%%%%%%%%%%%%%%%%%%%%%
\begin{table}[htb]
\caption{\label{tab:table1}A comparison of the known B(E2) values for the E2 transitions between yrast states in $^{130}$Cd with the LSSM calculations using the “$jj45pna$” and modified “$jj45pnamo$” interactions.}
\begin{ruledtabular}
\begin{tabular}{c c c c}
\multicolumn{4}{c}{B(E2) values (in units of $e^2 fm^4$)}\\
\hline
E2 transitions   &  Expt. &  $jj45pna$ & $jj45pnamo$ \\
\hline
\hline
&&&\\
${2^+ \rightarrow 0^+}$ & --	& 109.1 & 133.9 \\
${4^+ \rightarrow 2^+}$ & --	& 151.8	& 163.2 \\
${6^+ \rightarrow 4^+}$ & --	& 113.5	& 114.5 \\
${8^+ \rightarrow 6^+}$ & 50	& 46.0	& 46.0 \\
\end{tabular}
\end{ruledtabular}
\end{table}
%%%%%%%%%%%%%%%%%%%%%%%%%%%%%%%%%%%%%%%%%%%%%%%%%%%%%%%%%%%%%%%%%%%%%%%%%%%%%%%%%%%%%%%%%%%%%%
	
We plot in Fig. 2, the calculated average occupancies of the valence protons in $f_{5/2}$, $p_{3/2}$, $p_{1/2}$ and $g_{9/2}$ orbits from $0^+$ to $8^+$ states with “$jj45pna$” and “$jj45pnamo$” interactions. The purity of $\Pi g_{9/2}^{-2}$ configuration increases with the modified interaction, as the $g_{9/2}$ occupancy shifts towards $8$ particles for the $0^+$ to $4^+$ states. However, it remains nearly pure $8$ particles, or $2$ holes $\Pi g_{9/2}^{-2}$ configuration for the higher spin $6^+$ and $8^+$ states irrespective of the modifications. To further check this result, we plot in Fig.~\ref{fig:fig3} the percentage domination of $\Pi g_{9/2}^{-2}$ configuration from $0^+$ to $8^+$ states for the wave function in $^{130}$Cd, with “$jj45pna$” and “$jj45pnamo$” interactions. The dominance of $\Pi g_{9/2}^{-2}$ in the total wave function is too obvious, strengthening the pure seniority nature of these states.

As a consequence, the B(E2) values for the $2^+ \rightarrow 0^+$ and $4^+ \rightarrow 2^+$ transitions change marginally after modifying the interaction, as shown in Table~\ref{tab:table1}. The B(E2; $6^+ \rightarrow 4^+$) value is least affected by the modification, while the B(E2; $8^+ \rightarrow 6^+$) value and corresponding gamma ray energy for the $8^+$ isomer remains the same ($46$ $e^2fm^4$ and $145$ keV, respectively) with the modified “$jj45pnamo$” interaction. This may be attributed to the purity of the seniority quantum number and the maximally aligned spins of the two $g_{9/2}$ proton holes in the $8^+$ isomer. Because the $\langle \Pi g_{9/2}^{-2} \mid V \mid  \Pi g_{9/2}^{-2} \rangle $ TBME are enhanced by a constant factor in the modified interaction, the character of the $6^+$ and $8^+$ states remains unchanged, confirming the pure $g_{9/2}^{-2}$ nature of these states. However, the mixing of p-orbits is quite evident in the ground $0^+$ and the first excited $2^+$ states in original interaction, as seen from Figs.~\ref{fig:fig2} and ~\ref{fig:fig3}. Therefore, the corresponding E2 transition matrix elements change when the modified interaction is used since the relative gap of $p-$orbits with $g_{9/2}$ orbit changes significantly. No data for the B(E2) values corresponding to the E2 transitions ($2^+ \rightarrow 0^+$ , $4^+ \rightarrow 2^+$ and $6^+ \rightarrow 4^+$) are available~\cite{ensdf} and must be measured in future experiments. However, a configuration mixing in the lower lying states such as $0^+$ and $2^+$ states can be observed in Fig.~\ref{fig:fig2}.

The level energies come closer to the experimental data when we modify the interaction, which implies that the levels are mainly composed of the seniority $v=2$, $\Pi g_{9/2}^{-2}$ configuration in $^{130}$Cd. Since no core excitations have been used in these calculations, $N=82$ core breaking is not needed to explain the seniority isomerism. The origin of the meta-stable states can be understood in terms of pure seniority and maximally aligned states of a given-j orbit. Besides, the $8^+$ isomer in $^{128}$Pd (experimentally the last known even-even $N=82$ case before $^{132}$Sn) may also be understood in a similar manner. Half-life of the $8^+$ isomer in $^{128}$Pd is larger than $^{130}$Cd~\cite{jungclaus2007}, as expected from the seniority scheme. The seniority scheme predicts the $8^+$ isomer in lighter isotopes till $Z=42$ (with $N=82$), which may become accessible in future. However, the possibilities of seniority mixing cannot be ruled out, as in the case of $^{72,74}$Ni~\cite{isacker2011}. Further measurements are needed to confirm these possibilities. Also, the melting of the $N=82$ shell closure below $^{132}$Sn can be addressed decisively only when lighter waiting-point nuclei become accessible in future.

It would be interesting to check the role of modifications made in TBME for reproducing the level energies and B(E2) values in a consistent way. We, therefore, calculate the effective single particle energies (ESPE) for the active proton orbits $f_{5/2}$, $p_{3/2}$, $p_{1/2}$ and $g_{9/2}$ by using the formula~\cite{otsuka2010}:
\begin{eqnarray}
E_{espe} = E_{j_p}+ \Sigma_{j_p^\prime} E_{avg}(j_p,j_p^\prime) n_{j_p^\prime}
\end{eqnarray}
where $E_{j_p}$ an $n_{j_p^\prime}$ denote the single-particle energies and number of protons occupying the $j_p$ and $j_p^\prime$ orbits, respectively. The term $E_{avg} (j_p, j_p^\prime)$ represents the angular momentum averaged energy contributing to the monopole shift in energy as $\sum_{j_p^\prime} E_{avg}(j_p,j_p^\prime) n_{j_p^\prime}$ and may be obtained as:

\begin{eqnarray}
E_{avg} (j_p,j_p^\prime)= \frac {\sum_J (2J+1) \langle j_p j_p^\prime; J \mid V \mid j_p j_p^\prime; J \rangle } {\sum_J (2J+1)}
\end{eqnarray}
where $ \langle j_p j_p^\prime; J \mid V \mid j_p j_p^\prime; J \rangle $ represents the TBME taken from the shell model interaction. Near the shell closures, the monopole component affects according to $\sum_{j_p^\prime} E_{avg}(j_p,j_p^\prime) n_{j_p^\prime}$, and the effects of other multipole components vanish. The values at the valence proton number $=0$ show the single-particle energies of the proton orbits as given by the effective interaction. Note that the TBME involved in the monopole shift in proton orbits belong to $T=1$ components due to identical nucleons. One can find a certain gap between the $g_{9/2}$($jj45pna$) and p-orbits till the valence proton number $=16$. They come closer after $16$ and finally get mixed with each other for valence proton numbers $= 20-22$. As soon as we modify the interaction, the gap between the $g_{9/2}$($jj45pnamo$) and $p-$orbits begins to vanish linearly on going from valence proton number $=6$ to $16$. The $g_{9/2}$($jj45pnamo$) then crosses the p-orbits which eventually leads to nearly pure $g_{9/2}$ configuration in $^{130}$Cd, particularly for high-spin states. 

We find that the $g_{9/2}$ orbit (with $jj45pnamo$ interaction) lies at the Fermi surface near the valence proton number $20$ in $^{130}$Cd, and a significant gap develops between the $g_{9/2}$($jj45pnamo$) and the lower lying $p-$orbits. The $100$ keV increment in $g_{9/2}$ TBME actually results in a significant monopole correction. Therefore, the change in the location of the $g_{9/2}$ orbit as obtained from the $jj45pna$ and $jj45pnamo$ interactions becomes quite large; see Fig.~\ref{fig:fig4}. Our analysis, however, confirms the pure seniority nature of the $8^+$ isomer in $^{130}$Cd in both the interactions. This further suggests that the nature of effective interaction is conserved in the seniority scheme. Besides, ESPE maintain the $g_{9/2}$ orbit at the Fermi surface till valence proton number $=18$ that is for $^{128}$Pd. However, a significant configuration mixing between $g_{9/2}$ and p-orbits is expected while going towards the neutron-rich light mass nuclei of this region. This may then require a generalized seniority description, similar to the previous case of neutron-rich Sn isotopes beyond $^{132}$Sn~\cite{jain20171}.

\subsection{$^{206}$Hg}

\begin{figure*}[!ht]
\includegraphics[width=16cm,height=14cm]{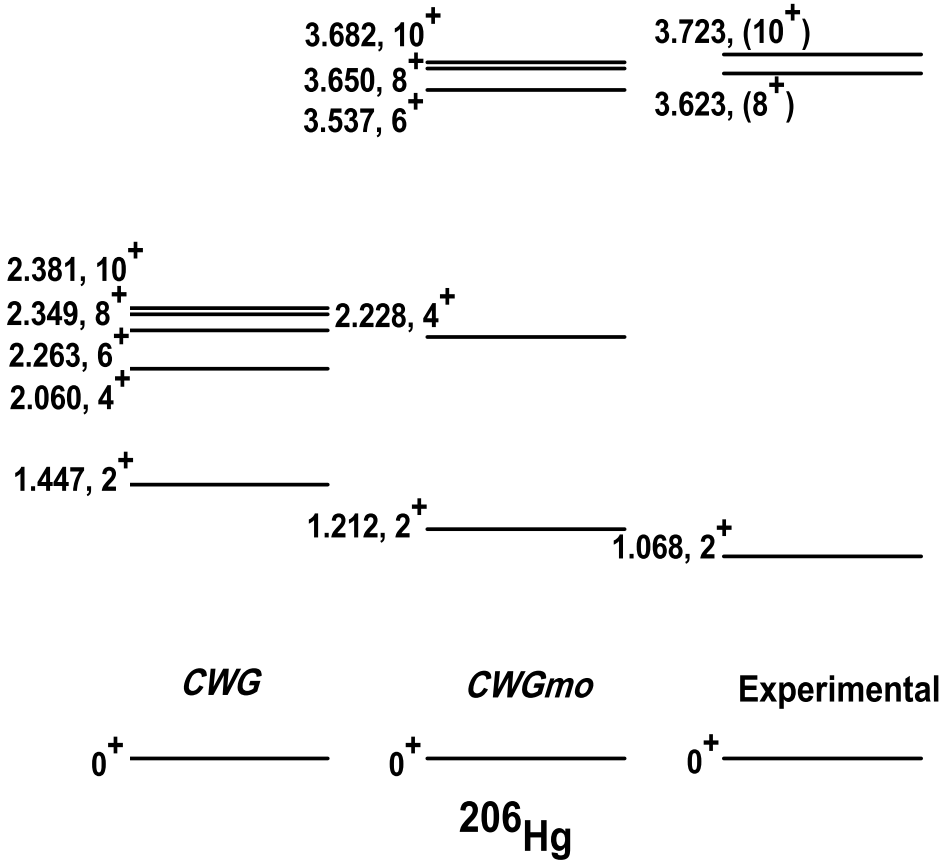}% Here is how to import EPS art
\caption{\label{fig:fig5} The calculated level schemes from $0^+$ to ${10}^+$yrast states in $^{206}$Hg with “$CWG$” and modified “$CWGmo$” interactions, along with the experimental data ~\cite{ensdf}. The calculated scheme with modified interaction comes closer to the experimental data. The experimental $4^+$ and $6^+$ states are not known, and the spin-parity assignments to $8^+$ and ${10}^+$ states are tentative in nature~\cite{ensdf}.}
\end{figure*}

We plot in Fig.~\ref{fig:fig5}, the experimental and the calculated level schemes of $^{206}$Hg with the “$CWG$” and modified “$CWGmo$” interactions. We see that the $2^+$ state is overestimated by $379$ keV, while the $8^+$ and ${10}^+$ states are underestimated by more than $1$ MeV by the “$CWG$" interaction. The $4^+$ and $6^+$ states are not known experimentally. The calculated ratio $E(4^+) / E(2^+)$ of $1.423$ again suggests a single-particle nature. The calculated E2 decay energy for the ${10}^+$ state is estimated to be small as compared to the experimental value ($100$ keV). The experimental~\cite{ensdf} and “$CWG$” calculated B(E2) values for the ${10}^+$ isomer are nearly $72$ and $62$ $e^2fm^4$, respectively.

\begin{figure*}[!ht]
\includegraphics[width=16cm,height=12cm]{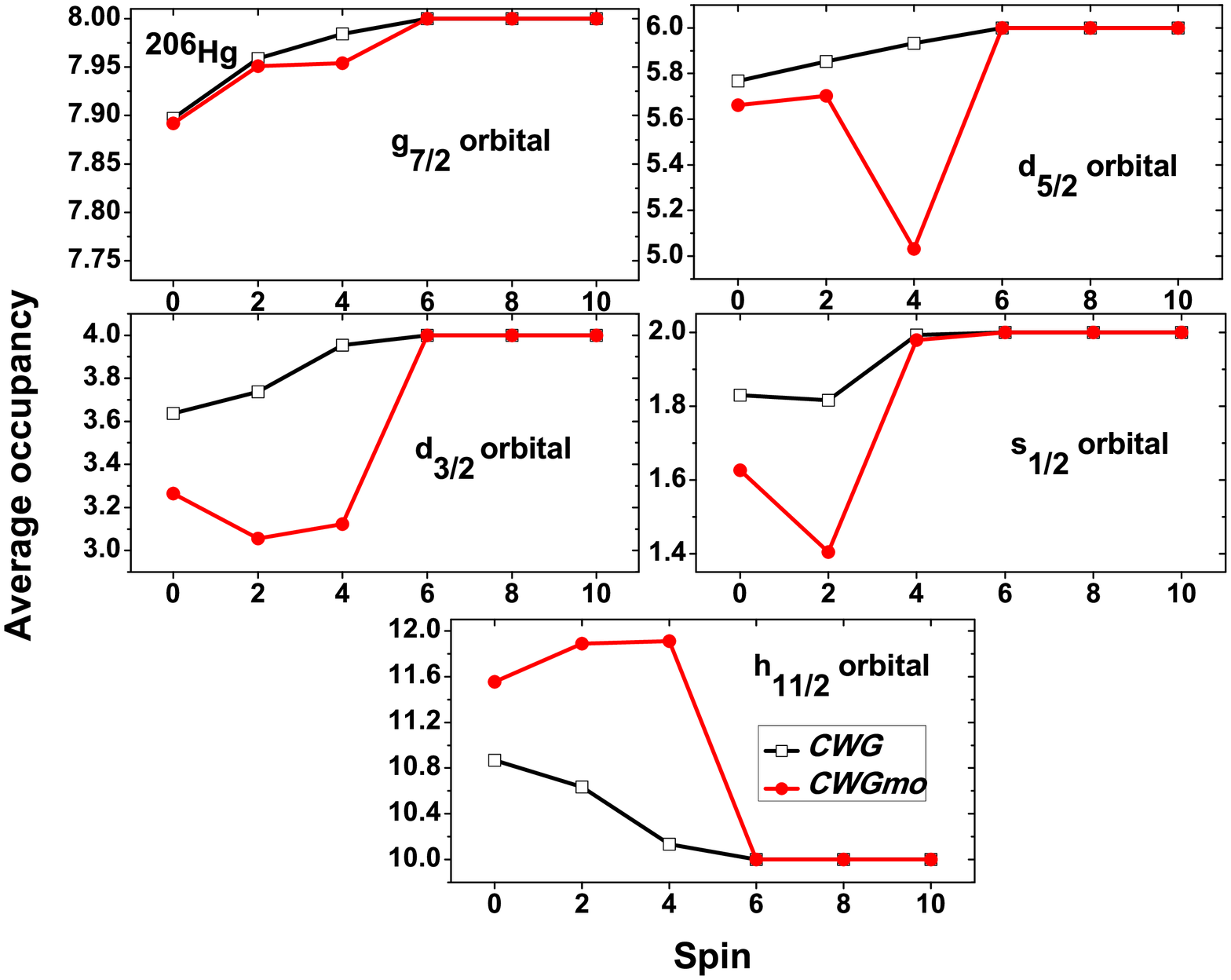}% Here is how to import EPS art
\caption{\label{fig:fig6}(Color online) The calculated average occupancies of the valence protons in $g_{7/2}$, $d_{5/2}$, $d_{3/2}$, $s_{1/2}$ and $h_{11/2}$ orbits from $0^+$ to ${10}^+$ spins for $^{206}$Hg with “$CWG$” and “$CWGmo$” interactions. These occupancies show the dominance of $\Pi h_{11/2}^{-2}$ configuration for the $6^+$, $8^+$ and ${10}^+$ states which persists with the modified interaction. The $0^+$ to $4^+$ states exhibit a significant mixing of $d-s$ orbits, particularly in the $4^+$ state when $CWGmo$ interaction is used.}
\end{figure*}

\begin{figure}[!ht]
\includegraphics[width=9.5cm,height=9cm]{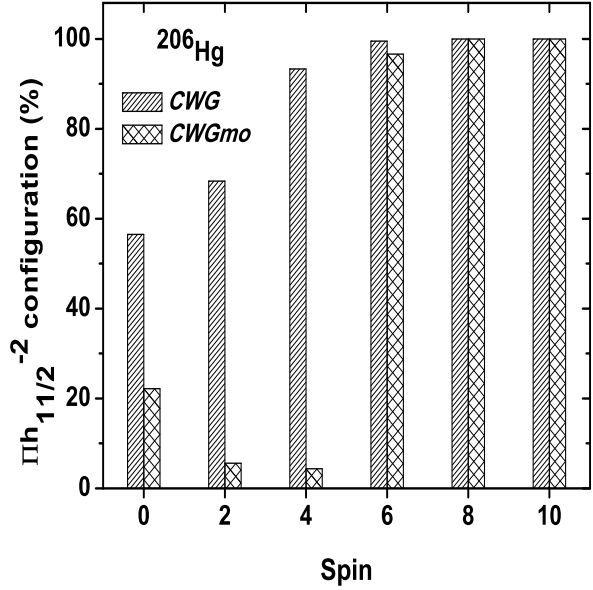}% Here is how to import EPS art
\caption{\label{fig:fig7}(Color online) The calculated percentage of $\Pi h_{11/2}^{-2}$ configuration in the resulting wave function of the valence protons from $0^+$ to ${10}^+$ spins for $^{206}$Hg with the “$CWG$” and the modified “$CWGmo$” interactions. One can follow the abrupt change in the wave functions for the $0^+$ to $4^+$ states after modifying the interaction while slightly change for $6^+$ state and no change at all for the $8^+$ and ${10}^+$ states.}
\end{figure}

\begin{figure}[!ht]
\includegraphics[width=9.5cm,height=9cm]{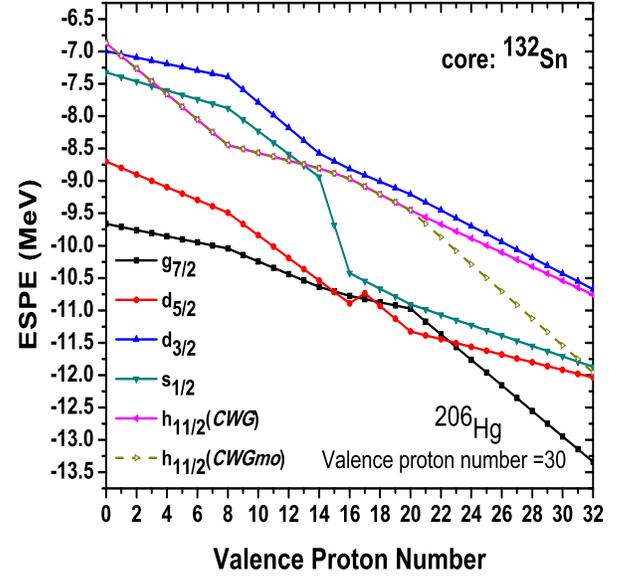}% Here is how to import EPS art
\caption{\label{fig:fig8}(Color online) The variation of ESPE for the active proton orbits $g_{7/2}$, $d_{5/2}$, $d_{3/2}$, $s_{1/2}$, and $h_{11/2}$. The ESPE of $h_{11/2}$ have been shown with the “$CWG$” and the modified “$CWGmo$” interactions. One can observe the large monopole shift at valence proton number $=30$ ($^{206}$Hg) for $h_{11/2}$ after a marginal change of 100 keV in the diagonal $\Pi h_{11/2}^{-2}$ TBME.}
\end{figure}

\begin{figure}[!ht]
\includegraphics[width=9.5cm,height=9cm]{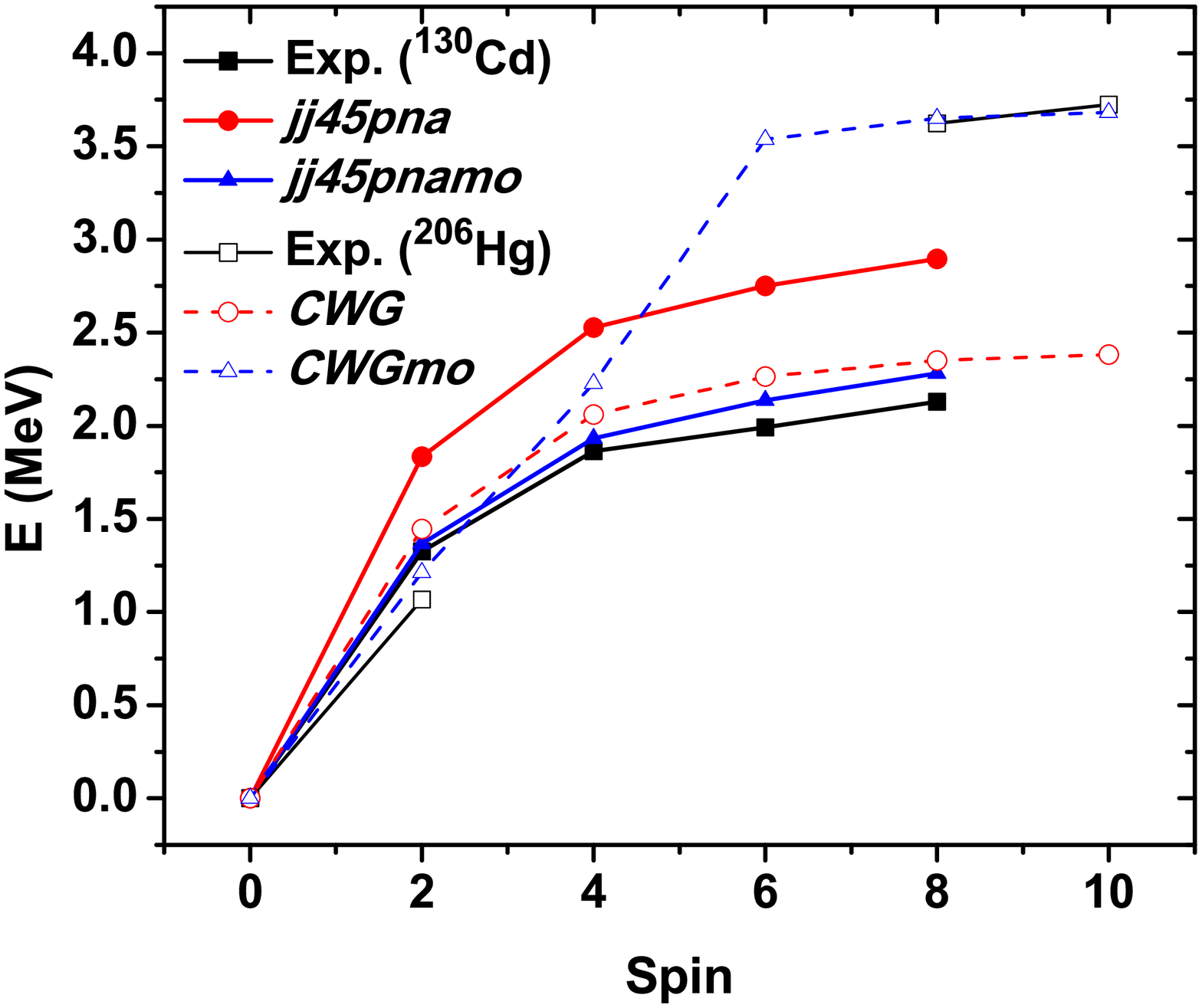}% Here is how to import EPS art
\caption{\label{fig:fig9}(Color online) The variation of experimental ~\cite{ensdf} and calculated energies versus spin of the nuclear states, for both $^{130}$Cd and $^{206}$Hg. See the discussion for more details. }
\end{figure}

However, the reduction in the TBME of $\Pi h_{11/2}^{-2}$ by $100$ keV brings up the calculated level energies closer to the experimental data as shown in Fig.~\ref{fig:fig5}. We call this modified interaction as “$CWGmo$”, where only $6$ diagonal $\Pi h_{11/2}^{-2}$ TBME $ \langle \Pi h_{11/2}^{-2} \mid V \mid \Pi h_{11/2}^{-2} \rangle$, corresponding to $0^+$, $2^+$, $4^+$, $6^+$, $8^+$ and ${10}^+$ spins have been modified by reducing the TBME of $\Pi h_{11/2}^{-2}$ by $100$ keV. No core excitations have been taken into account for these shell model calculations. As a consequence, the B(E2) value for the ${10}^+$ isomer remains unaltered after the modification as there is no scope of seniority mixing, similar to the case of $8^+$ isomer in $^{130}$Cd.

%%%%%%%%%%%%%%%%%%%%%%%%%%%%%%%%%%%%%%%%%%%%%%%%%%%%%%%%%%%%%%%%%%%%%%%%%%%%%%%%%%%%%%%%%%%%
\begin{table}[htb]
\caption{\label{tab:table2}The experimental and shell model calculated B(E2) values for the E2 transitions between yrast states in $^{206}$Hg with “$CWG$” and modified “$CWGmo$” interactions.}
\begin{ruledtabular}
\begin{tabular}{c c c c}
\multicolumn{4}{c}{B(E2) values (in units of $e^2 fm^4$)}\\
\hline
E2 transitions   &  Expt. &  $CWG$ & $CWGmo$ \\
\hline
\hline
&&&\\
${2^+ \rightarrow 0^+}$ & -- & 336.0 & 416.5 \\
${4^+ \rightarrow 2^+}$ &--	& 324.1	& 479.9 \\
${6^+ \rightarrow 4^+}$ & -- & 	261.3 &	38.1 \\
${8^+ \rightarrow 6^+}$ & -- &	166.6 &	161.7 \\
${{10}^+ \rightarrow 8^+}$ & 72 & 62.7 & 62.7 \\
\end{tabular}
\end{ruledtabular}
\end{table}
%%%%%%%%%%%%%%%%%%%%%%%%%%%%%%%%%%%%%%%%%%%%%%%%%%%%%%%%%%%%%%%%%%%%%%%%%%%%%%%%%%%%%%%%%%%%%%
	
The B(E2) values for the $2^+ \rightarrow 0^+$ and $4^+ \rightarrow 2^+$ transitions change significantly after modifying the interaction as shown in Table~\ref{tab:table2}. The B(E2; $8^+ \rightarrow 6^+$) value gets slightly changed, while the B(E2; ${10}^+ \rightarrow 8^+$) and corresponding E2 decay energy for the ${10}^+$ isomer remains the same with “$CWGmo$”. An exception is the B(E2; $6^+ \rightarrow 4^+$) value which reduces abruptly with the “$CWGmo$” interaction. This may be understood by the presence of two low lying $4^+$ states below the yrast $6^+$ state when “$CWGmo$” is used. We conclude that the yrast $6^+$ and $4^+$ states acquire different seniorities when the “$CWGmo$” interaction is used. For the second $4_2^+$ state, we get B(E2; $6^+ \rightarrow 4_2^+$) value as $206.1$ $e^2fm^4$(from “$CWGmo$”), which is of the same order as the “ $CWG$” calculated B(E2; $6^+ \rightarrow 4^+$) value of $261.3$ $e^2fm^4$ as shown in Table~\ref{tab:table2}. Both the $4^+$ and $4_2^+$ states have different configurations and the second $4_2^+$ state is the one dominated by the $\Pi h_{11/2}^{-2}$ configuration when “$CWGmo$” interaction is used.

If we analyze the average occupancies of the active proton orbits in $^{206}$Hg, as shown in Fig.~\ref{fig:fig6}, the dominance of $\Pi h_{11/2}^{-2}$ configuration is quite visible from $6^+$ to ${10}^+$ states in both the original and the modified interactions. A significant difference is, however, seen for the lower lying $0^+$ to $4^+$ states. The $0^+$ to $4^+$ states avoid the $\Pi h_{11/2}^{-2}$ configuration and have the dominance of $d-s$ orbits in their respective configurations. This can also be followed from Fig.~\ref{fig:fig7}, where we have plotted the calculated percentage of $\Pi h_{11/2}^{-2}$ configuration in the resulting wave functions. The percentage reduces to less than five for the yrast $2^+$ and $4^+$ states with “$CWGmo$” interaction leading to a significant change in the picture before and after reducing the $\Pi h_{11/2}^{-2}$ TBME.  

However, the ${10}^+$ isomer remains untouched in the “$CWGmo$” calculations. This isomer decays to the low lying $8^+$ state via E2 transition and there is no chance of any seniority mixing in this proton valence space. So, the B(E2; ${10}^+ \rightarrow 8^+$) value remains unaltered even after this modification as no change in the E2 transition matrix elements could be made. These LSSM calculations do not allow any particle-hole excitations across $Z=82$ and $N=126$. This means that no shell quenching is needed to explain the maximally aligned isomer having a pre-dominance of $\Pi h_{11/2}^{-2}$ configuration. We note that higher spin states (with tentative spin-parity assignments) like ${12}^+$ and ${13}^-$ having energy more than $5.5$ MeV are also known in $^{206}$Hg with tentative spin-parity assignments. These cannot be explained by seniority $v=2$, $\Pi h_{11/2}^{-2}$ configuration and the present LSSM calculations. The core excitations across $Z=82$ or $N=126$ shell may then become necessary to explain the origin of such states.  

Results are quite interesting as the small changes in the limited TBME of the effective interaction can significantly alter the location of some orbits, the wave functions and the corresponding spectral features. For example, the single-particle energies of the “$CWG$” interaction have been taken to reproduce the mass and spectra of $^{133}$Sn (odd-neutron, $N=83$) and $^{133}$Sb (odd-proton, $Z=51$)~\cite{brown2005}, which seem to be inadequate for $^{206}$Hg ($Z=80$, $N=126$) having $2-$proton holes. More rigorous theoretical efforts are indeed required to resolve such issues where the single-particle energies taken from a shell closure along with the related matrix elements may not remain applicable while approaching another shell closure. 

To investigate it further, we have plotted in Fig.~\ref{fig:fig8}, the ESPE for the active proton orbits $g_{7/2}$, $d_{5/2}$, $d_{3/2}$, $s_{1/2}$, and $h_{11/2}$, by using Equation (1). Note that the TBME involved in the monopole shift in proton orbits belong to $T=1$ components due to identical nucleons. We can see that the $h_{11/2}$, along with $d_{3/2}$ orbit, lie at the top (refer to the single-particle energies at valence proton number $=0$ in Fig.~\ref{fig:fig8}). As soon as the protons begin to fill in this active valence space of $50-82$, the monopole shift results in the shifting of $h_{11/2}$($CWG$) towards lower lying $g_{7/2}$ and $d_{5/2}$ orbits till valence proton number $=14$; however, a certain gap is maintained. After the middle (valence proton number $=16$), the $s_{1/2}$ suddenly shifts towards lower lying $g_{7/2}$ and $d_{5/2}$, while $d_{3/2}$ and $h_{11/2}$($CWG$) remain close to each other. On modifying the interaction by $100$ keV, the $h_{11/2}$($CWGmo$) shifts lower than $h_{11/2}$($CWG$). The shift is monotonically increasing as we approach the magic number $Z=82$. Interestingly, the modified behavior of $h_{11/2}$($CWGmo$) orbit is quite similar to that of $g_{7/2}$ orbit. At valence proton number $= 30$ in $^{206}$Hg, the gap between $d_{3/2}$ and $h_{11/2}$($CWGmo$) becomes quite large, and $d_{3/2}$ becomes active near the Fermi surface, leading to a mixing of $d_{3/2}$ orbit in low-spin states, also seen in the LSSM results. However, the purity of high-spin states remains untouched due to the seniority $v = 2$ configuration states in two-proton hole Hg nucleus. This may be understood from the obvious generation of ${10}^+$ and $8^+$ states from $h_{11/2}$ orbital, as the $s-d$ orbits cannot contribute in their resulting wave functions, particularly when only two holes are present.

Interestingly, the half-life of the ${10}^+$ isomer (with a tentative spin-parity assignment) in $^{204}$Pt is larger than the half-life in $^{206}$Hg, as expected in the seniority scheme, which predicts a parabola for the B(E2)s and inverted parabola for the half-lives with respect to nucleon number. However, a configuration mixing is expected for lighter mass $^{204}$Pt where four holes are involved. It may then become necessary to invoke the generalized seniority description; which is beyond the scope of the present paper. 

We have plotted in Fig.~\ref{fig:fig9}, the experimental and calculated energies of the lowest excited states versus their spins and compared them with calculated results for both the modified and the unmodified interactions. All the nonzero spin states are seniority $v=2$ states except for the second $4^+$ state in the “$CWGmo$” results for $^{206}$Hg, which may have a different seniority, as suggested by the B(E2) values shown in Table~\ref{tab:table2}. We also notice an unusual steep rise in the “$CWGmo$” values due to this. We also recall that the shell model analysis does not support any configuration mixing in both $^{130}$Cd and $^{206}$Hg and the seniority remains constant at seniority $v=2$. 

The most surprising finding from Fig.~\ref{fig:fig9} is the closeness of the energies from the “$CWG$” calculations for $^{206}$Hg with the experimental data for $^{130}$Cd and also with the “$jj45pnamo$” energy curve for $^{130}$Cd. We believe that the single-particle energies used in “$CWG$” are responsible for this as these are taken to reproduce the latest mass data and spectra of $^{133}$Sb and $^{133}$Sn~\cite{jones2010}. Fig.~\ref{fig:fig9} brings out the limitations of this interaction when dealing with the nuclei near $Z=82$, particularly $^{206}$Hg in the present work. It highlights the need to improve the interaction by using the data from another shell closure  $^{207}$Tl ($Z=81$, $N=126$)~\cite{ensdf}.  

\section{\label{sec:level4} Summary and Conclusion}

A parallel between the $8^+$ isomer in $^{130}$Cd and the ${10}^+$ isomer in $^{206}$Hg has been used as a stepping stone to check the robustness of the respective $N=82$ and $N=126$ magic numbers in neutron-rich region, particularly when very limited data are available in these waiting-point nuclei. We have presented a comparative shell model study of the $8^+$ isomer in $^{130}$Cd and ${10}^+$ isomer in $^{206}$Hg, which are shown to be pure seniority $v=2$ isomers with maximally aligned spin arising from the intruder orbits in the respective valence spaces. These theoretical efforts also shed light on the spin-parity assignments, and related structural changes around doubly magic configurations in these neutron-rich nuclei, which are very important in astrophysical calculations. However, modifications in the shell model interactions are required to consistently match the level schemes and the B(E2) values. The limitations in shell model interactions may be improved by using the consistent set of single-particle energies. The seniority scheme also predicts $8^+$ isomer in lighter isotopes $Z<46$ (with $N=82$), and ${10}^+$ isomer in $Z<78$ (with $N=126$) which may become accessible in future.

We explain the similar behavior of isomers while going from a semi-magic nucleus $^{130}$Cd ($Z=48$, $N=82$) to the other semi-magic nucleus $^{206}$Hg ($Z=48$, $N=126$) in terms of seniority and confirm the robustness of the $N=82$ and $126$ magic numbers for these meta-stable states. These studies do not support any shell quenching and support the usual shell model gaps in the single-particle structure.

\begin{acknowledgments}

BM gratefully acknowledges the financial support in the form of Post-Doctoral Research Fellowship from University of Malaya. HAK and NY acknowledge support from the Research University Grant (GP0448-2018) under University of Malaya. AKJ thanks Amity Institute of Nuclear Science and Technology, Amity University UP, for the financial support to continue this work.  
\end{acknowledgments}

\newpage %Just because of unusual number of tables stacked at end
\bibliography{apssamp}% Produces the bibliography via BibTeX.

\end{document}